\documentclass[preprint]{ptephy_v1}
\usepackage[latin1]{inputenc}
\usepackage[T1]{fontenc}
\usepackage[english]{babel}
\usepackage{amsmath}
\usepackage{amssymb,amsfonts,textcomp}
\usepackage{color}
\usepackage{array}
\usepackage{supertabular}
\usepackage{hhline}
\usepackage{hyperref}
\hypersetup{pdftex, colorlinks=true, linkcolor=blue, citecolor=blue, filecolor=blue, urlcolor=blue, pdftitle=, pdfauthor=Goh Eong Sheng, pdfsubject=, pdfkeywords=}
\makeatletter
\newcommand\arraybslash{\let\\\@arraycr}
\makeatother
\makeatletter
\newcommand\ps@Standard{
  \renewcommand\@oddhead{}
  \renewcommand\@evenhead{}
  \renewcommand\@oddfoot{}
  \renewcommand\@evenfoot{\@oddfoot}
  \renewcommand\thepage{\arabic{page}}
}
\newcommand\ps@FirstPage{
  \renewcommand\@oddhead{}
  \renewcommand\@evenhead{}
  \renewcommand\@oddfoot{*Corresponding author}
  \renewcommand\@evenfoot{\@oddfoot}
  \renewcommand\thepage{\arabic{page}}
}
\makeatother
\renewcommand\arraystretch{1.3}
\newcounter{Table}

\newcounter{Figure}

\providecommand{\keywords}[1]{\textbf{\textit{Index terms---}} #1}

\begin{document}

\title{Effects of Hubbard term correction on the structural parameters and
electronic properties of wurtzite ZnO}

\author{E. S. Goh}
\affil{School of Physics, Universiti Sains Malaysia, 11800 USM Penang, Malaysia
\email{sheng5931@hotmail.com }}

\author{J. W. Mah}
\affil{School of Physics, Universiti Sains Malaysia, 11800 Penang, Malaysia
\email{jianwen$\_$mah@yahoo.com}
}


\author{T. L. Yoon}
\affil{School of Physics, Universiti Sains Malaysia, 11800 Penang, Malaysia
\email{tlyoon@usm.my}}

\begin{abstract}
The effects of including the Hubbard on-site Coulombic correction to the structural parameters and valence energy states of wurtzite ZnO was
explored. Due to the changes in the structural parameters caused by
correction of hybridization between Zn d states and O p states,
suitable parameters of Hubbard terms have to be determined for an
accurate prediction of ZnO properties. Using the LDA+${U}$ method
by applying Hubbard corrections $U_p$ to
Zn 3d states and $U_p$ to O 2p states, the lattice constants were underestimated for all tested Hubbard parameters. The combination of both $U_d$ and $U_p$ correction terms managed to widen the band gap
of wurtzite ZnO to the experimental value. Pairs of $U_p$ and
$U_p$ parameters with the correct positioning of d-band and accurate bandwidths were selected, in addition to predicting an accurate band gap value. Inspection of vibrational properties, however, revealed mismatches between the estimated gamma phonon frequencies and experimental values. The selection of Hubbard terms based on electronic band properties alone cannot ensure an accurate vibrational description in LDA+${U}$ calculation.
\end{abstract}



\maketitle

\section{Introduction}

Zinc oxide had been known as a relatively low cost and readily
synthesized material. It has the properties of a polar semiconductor
with a wide band gap of 3.44 eV with potential applications in the
optoelectronic industries [1]. Numerous theoretical calculation on the
properties of ZnO had been carried out using density functional theory
(DFT). However, it is widely known that a standard DFT calculation
typically suffers from a band gap problem, where the band gap of a
material is grossly underestimated [2]. This could lead to an
inaccurate estimation of the electronic properties of a material such
as ZnO, which is a potential material for the optoelectronic industry.

The band gap problem of DFT calculation can be addressed by including
the GW approximation in which the self-energy of a many-body system of
electrons is taken into account [3]. Zhang \textit{et al.} estimated
that the band gap of wurtzite ZnO to be within the range of 2.82 eV to
4.54 eV by using various types of GW approximations [4]. An alternative
way to improve the band gap is to employ the hybrid exchange
correlation functionals in a DFT calculation. Using the HSE06
functional, Zhou predicted the band gap of wurtzite ZnO to be 2.79 eV
[5]. While both GW approximation and hybrid functional have proved to
be effective in mitigating the band gap in a standard DFT calculation,
a major downside is the requirement of a computational power much
higher than that required by a standard DFT computation. This has
limited the feasibility of conducting a realistic prediction of
properties of a material via DFT.

DFT+\textit{U} calculation has emerged as a means to improve the
electronic properties prediction at a computational cost comparable to
that required by a standard DFT calculation. In a DFT+\textit{U}
calculation, Hubbard-type interactions are included in the standard
exchange correlation functional of local density approximation (LDA) or
generalized-gradient approximation (GGA) through the Hubbard parameters
\textit{U} and \textit{J} [6, 7]. \ A method of estimating the values
for the Hubbard parameters \textit{U} and \textit{J} had been provided
by Cococcioni et al. through the linear response approach [8].
\ However, the linear response approach, which has been widely used for
the open shell system, is less than ideal in the case of a closed-shell
system such as ZnO with a full electronic shell of d orbitals.
Numerical reliability is an issue of concern in the case of completely
full localized bands which exhibit a very small response to linear
perturbation [9]. 

In this work, the dependence of structural parameters and electronic
properties of wurtzite ZnO on the different values of Hubbard
parameters \textit{U} and \textit{J} is investigated. A series of
DFT+\textit{U} calculation is carried out with different values of both
Hubbard parameters associated with Zn and O. While the Hubbard
parameters are introduced to improve electronic properties, the
structural parameters of a material are modified alongside with the
electronic properties; the selection of suitable values of Hubbard
parameters depends on the both structural and electronic properties. A
similar work on the LDA+\textit{U} and GGA+\textit{U} calculation of
wurtzite ZnO had been carried out by Huang \textit{et al.} [10] by
using the electronic calculation package VASP. However, different
results are obtained in this work through the use of different
pseudopotentials.

\section{Computational Methods}

The experimental structural parameters of wurtzite ZnO at
\textit{a}=3.2427 Å and \textit{c}= 5.1948 Å with wurtzite parameter
\textit{z}=0.3826 as discovered by Sabine and Hogg [11] using X-ray
crystallography method is used as the reference structural data in this
work. A standard DFT geometrical optimization is first performed on the
primitive wurtzite ZnO unit cell, followed by a self-consistent DFT
calculation to study its electronic properties. Hubbard term
$U_d$ is then added to the d-orbitals of
the Zn atoms, ranging from 2 eV to 14 eV in the interval of 2 eV. The
second part of this research studies the effect of including both
Hubbard term $U_d$ to Zn
d-orbitals and $U_p$ to O p-orbitals,
where the $U_p$ values ranges from 5 eV
to 9 eV. The changes to lattice constants and wurtzite parameter
\textit{z} as well as the band gap and valence band width are
investigated.

All calculations in this work are completed using the ABINIT
electronic package [12] within the framework of
projector-augmented-wave (PAW) potentials [13] and LDA exchange
correlation functionals. The PAW potentials used are from the datasets
provided by Jollet \textit{et al.} [14]. The plane wave basis sets are
expanded to kinetic energy cutoff of 34 Hartree whereas the
Monkhorst-Pack k-point mesh is set to an array of $8 \times 8 \times 6$
at gamma centred grid.

The LDA+\textit{U} calculations are performed using full localized
limit (FLL) double-counting correction [6]. The double-counting
correction is necessary in a LDA+\textit{U} calculation to avoid double
counting of correlation part in localized electrons. The Hubbard term
\textit{J} is set to zero for all calculations; the rotationally
invariant LDA+\textit{U} form proposed by Dudarev \textit{et al.} [7]
is equivalent to the FLL method with \textit{J}=0 and \textit{U} in
place of  $\bar{{U}}-\bar{{J}}$ [15]. 

\section{Results and Discussion}

\subsection{Standard DFT result}

Wurtzite ZnO possesses a space group number of 186 with hexagonal
symmetry and the following Wyckoff positions (see Table 1).
\begin{table}[h]
\label{table1}
\caption{Wyckoff positions of atoms in the primitive unit cell of wurtzite ZnO}
\begin{center}
\tablehead{}
\begin{supertabular}{m{0.49625984in}m{1.0427599in}m{2.9143599in}}
\hline
\centering Atom &
\centering Wyckoff letter &
\centering\arraybslash Wyckoff positions\\\hline
\centering Zn &
\centering 2b &
\centering\arraybslash (1/3,2/3,0), (2/3,1/3,1/2)\\
\centering O &
\centering 2b &
\centering \arraybslash (1/3, 2/3, \textit{z}),(2/3,1/3,
\textit{z}+1/2)\\\hline
\end{supertabular}
\end{center}
\end{table}
$z$ is the wurtzite parameter to be determined using DFT geometrical
optimization. The result obtained from a DFT geometrical optimization
followed by a self-consistent calculation is compared with the
corresponding experimental results in Table 2.
\begin{table}[h]
\label{table2}
\caption{Comparison between the experimental result of wurtzite ZnO and the
result from standard DFT LDA calculation.}
\begin{center}
\tablehead{}
\begin{supertabular}{m{0.8372598in}m{0.5677598in}m{0.5677598in}m{0.5677598in}m{0.5677598in}m{0.93935984in}m{1.0469599in}}
\hline
~
 &
\centering $a$ (Å) &
\centering $c$ (Å) &
\centering $c/a$ &
\centering $z$ &
\centering Volume (Å\textsuperscript{3}) &
\centering\arraybslash Band gap (eV)\\\hline
\centering Experiment\textsuperscript{*} &
\centering \textcolor{black}{3.2427} &
\centering 5.1948 &
\centering 1.6020 &
\centering 0.3826 &
\centering 47.3056 &
\centering\arraybslash 3.44\\
\centering LDA &
\centering \textcolor{black}{3.1836} &
\centering \textcolor{black}{5.1497} &
\centering \textcolor{black}{1.6175} &
\centering \textcolor{black}{0.3804} &
\centering 45.2022 &
\centering\arraybslash \textcolor{black}{0.7965}\\\hline
\end{supertabular}\\
{\footnotesize * The values for $a$, $c$, $c/a$ and $z$ are taken from Ref [11], whereas the band gap value is from Ref [16].}
\end{center}
\end{table}
The primitive unit cell volume is underestimated by 4.45\% compared to
the experimental volume, which is typical of a LDA calculation due to
overbinding in estimation for solids [17]. While both \textit{a} and
\textit{c} lattice constants are estimated to be lower than the
experimental values, the hexagonal unit cell is noticed to have
experienced elongation of approximately 1\% in the \textit{c} axis by
comparing the \textit{c/a} ratios. Regarding the atomic positions
within the unit cell, the heights of the two oxygen atoms are predicted
to be lower by about 0.0113 Å, due to the lower value of wurtzite
parameter \textit{z} found by LDA. On the other hand, the electronic
band gap value is predicted to be only approximately a quarter of the
experimental value, which is a serious underestimation known the
{\textquotedblleft}DFT band gap problem{\textquotedblright} as alluded
to earlier.

\subsection{Inclusion of Hubbard parameter $U_d$\textit{ to Zn}}

The dependence of the lattice constant \textit{a} and \textit{c} and
the \textit{c/a} ratio on $U_d$ is shown
in Figure 1. It shows that lattice constant \textit{a} increases
whereas the lattice constant \textit{c} decreases monotonically with
the increment in the value of Hubbard term
$U_d$ applied to the Zn d-orbitals.
However, the trend lines of the figures never reach the point of
experimental values, which are 3.2427 Å for \textit{a} and 5.1948 Å for
\textit{c}. While the values of lattice constant \textit{a} do increase
with $U_d$, the rate of increment is not
high enough to reach the experimental value even at
$U_d$=14 eV. The increasing trend of
lattice constant \textit{a} is also in contrast to the decreasing trend
found by Huang \textit{et al.} [10] in a similar study using VASP
within the PAW framework. Despite the mismatch of \textit{a} and
\textit{c} values, the \textit{c/a} ratio does approaches the
experimental value of 1.6020 between
$U_d$ values of 10 eV and 12 eV in its
decreasing trend. The wurtzite parameter \textit{z} increases with the
value of $U_d$, approaching but not
reaching the experimental value of 0.3826.

\begin{figure}
\label{fig1}
\begin{flushleft}
\tablehead{}
\begin{supertabular}{m{2.9761598in}m{3.16916in}}
{\centering 
\includegraphics[width=2.898in,height=2.2402in]{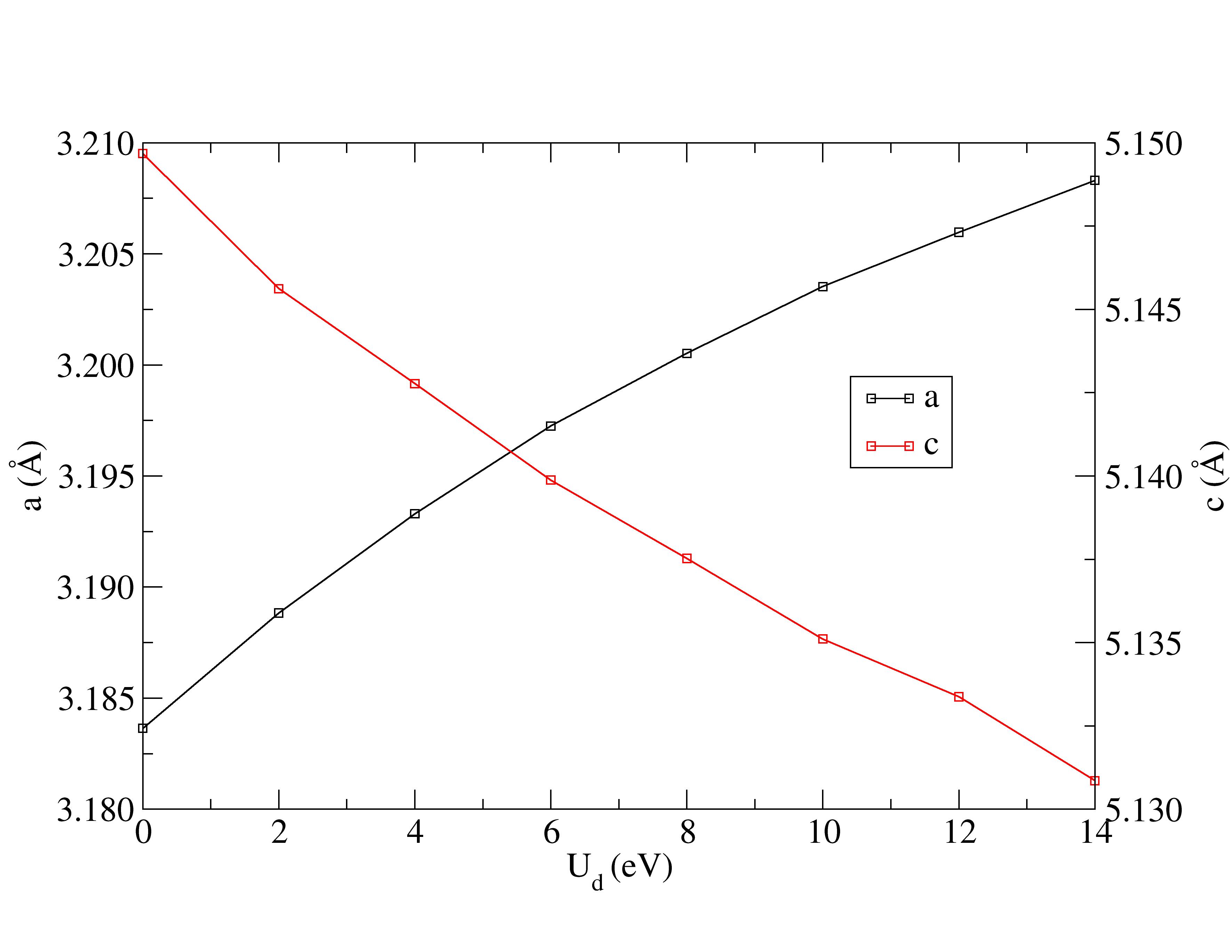} \par}
\centering (a) &
{\centering  \includegraphics[width=2.9in,height=2.2402in]{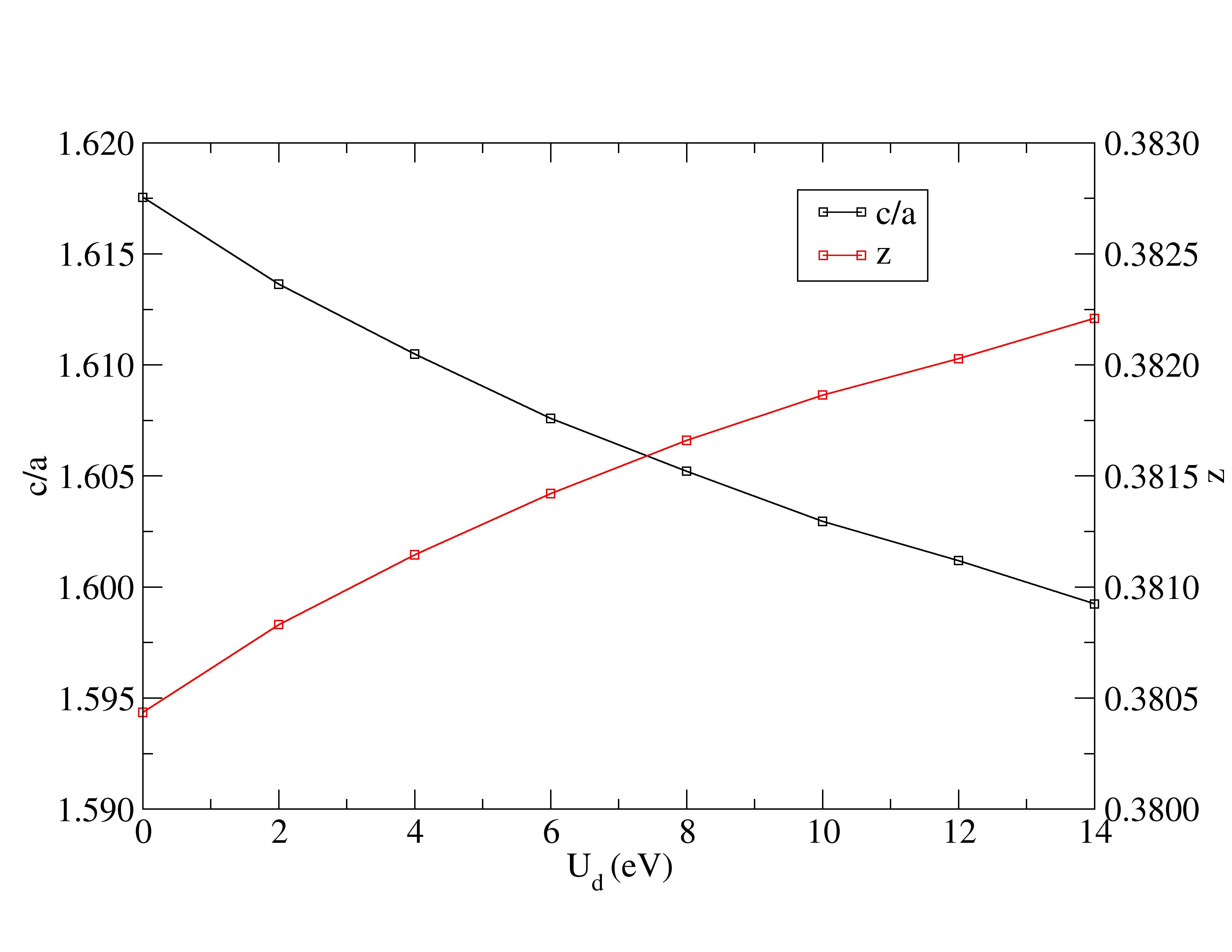}
\par}
\centering\arraybslash (b)\\
\end{supertabular}
\end{flushleft}
\caption{Variation of (a) lattice constant $a$ and $c$,(b) $c/a$ ratio and wurtzite parameter $z$ with respect to $U_d$.}
\end{figure}

The change in band gap with respect to
$U_d$ value is portrayed in Figure 2.
The band gap is shown to steadily increase to the highest value of 1.81
eV at $U_d$=14 eV. Despite the role of the Hubbard term
$U_d$ in interaction between localized
electrons, the estimated band gap is still too small with respect to
the ZnO experiment value of 3.44 eV at low temperature; the estimated
value is only approximately one half of the real value. Nevertheless,
our estimation of the structural parameters and band gap is in good
agreement with the LDA+\textit{U} results found by Kaczkowski [18].

\begin{figure}
\label{fig2}
{\centering 
\includegraphics[width=2.9035in,height=2.2437in]{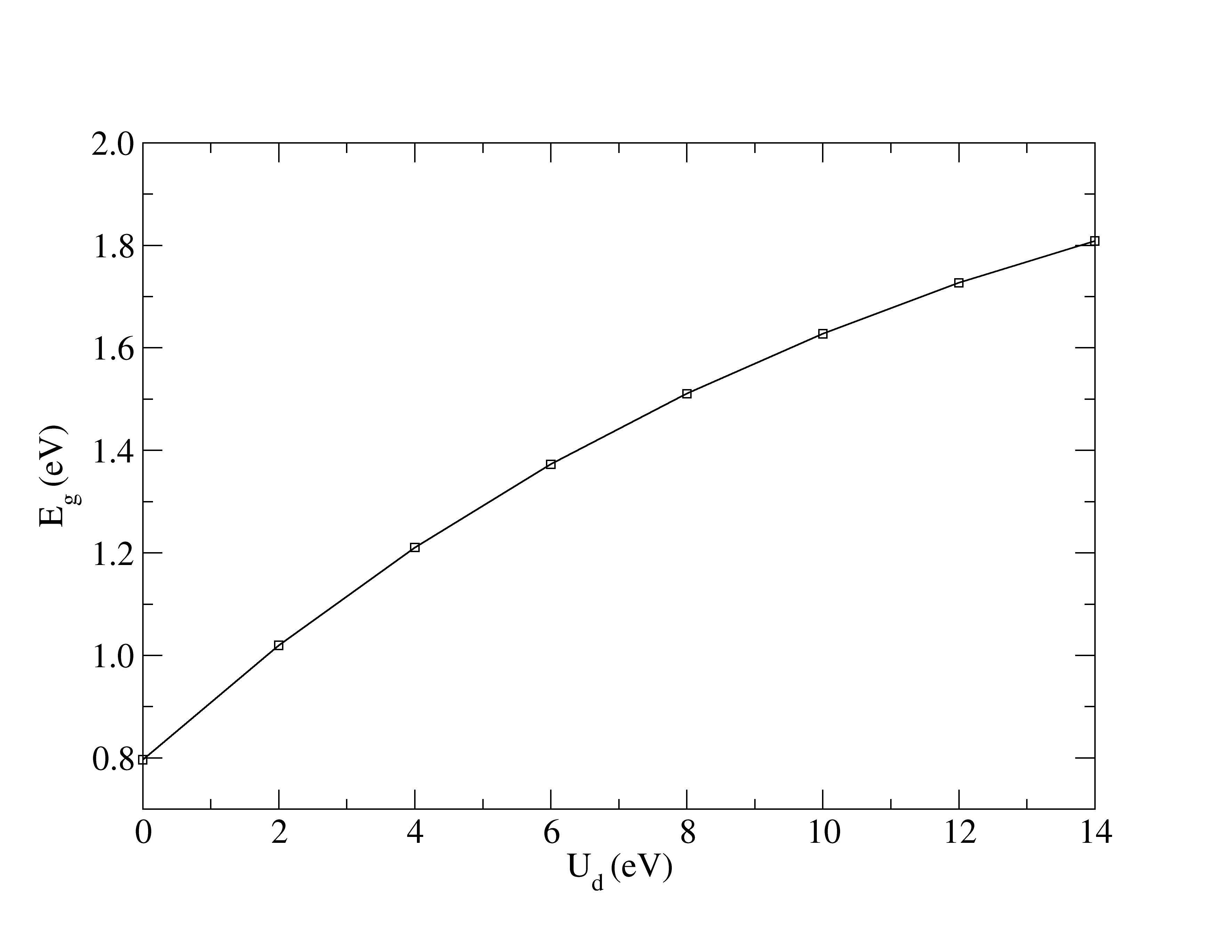} \par}
\caption{Variation of band gap with respect to Hubbard parameter $U_d$ applied to Zn d-orbitals}
\end{figure}

\subsection{Inclusion of Hubbard parameter $U_d$ to Zn and $U_p$ to O}

The dependence of the structural parameters upon the application of
on-site Coulomb correction $U_p$ on the O \textit{p}{}-orbitals is shown in Figure 3. In contrast to the
reduction of lattice constant \textit{a} with increasing
$U_d$, the lattice constant drops
monotonically with increasing $U_p$ at
all fixed values of $U_d$ due to
stronger localization of the electron states, in agreement with the
results produced by Huang \textit{et al.} [10] and Ma \textit{et al.}
[19]. The same trend is observed for lattice constant \textit{c}, in
which the height of the hexagonal unit cell deviates from the
experimental value of 5.1948 Å with increasing
$U_p$. On the contrary, the correlation
between \textit{c/a} ratio and estimated wurtzite parameter \textit{z}
with $U_p$ is weak; the changes in the
said parameters are insignificant. The reduction in the volume of unit
cell is revealed to be isotropic along both \textit{a} and \textit{c}
axis directions.

The influence of $U_p$ on the p-d
hybridization is evident by inspecting the band gap as shown in Figure
4. There is a huge enhancement in the magnitude of the electronic band
gap with the increment with value of
$U_p$. The rate of band gap widening is
high enough that the estimated band gaps have exceeded the
corresponding experimental value; the higher the
$U_d$ value, the lower the magnitude of
$U_p$ needed to widen the band gap to
match the experimental gap of 3.44 eV. Further investigations on the
electronic band properties are needed for the selection of suitable
values of $U_d$ and
$U_p$. Our result here is in contrast to
that found by Ma \textit{et al.} [19], in which the increment of
$U_d$ to 12.5 eV in GGA+U method is
sufficient to obtain a band gap that matches with that of experiment.

\begin{figure}[t]
\label{fig3}
\begin{flushleft}
\tablehead{}
\begin{supertabular}{m{2.9761598in}m{3.16916in}}
{\centering 
\includegraphics[width=2.7217in,height=2.102in]{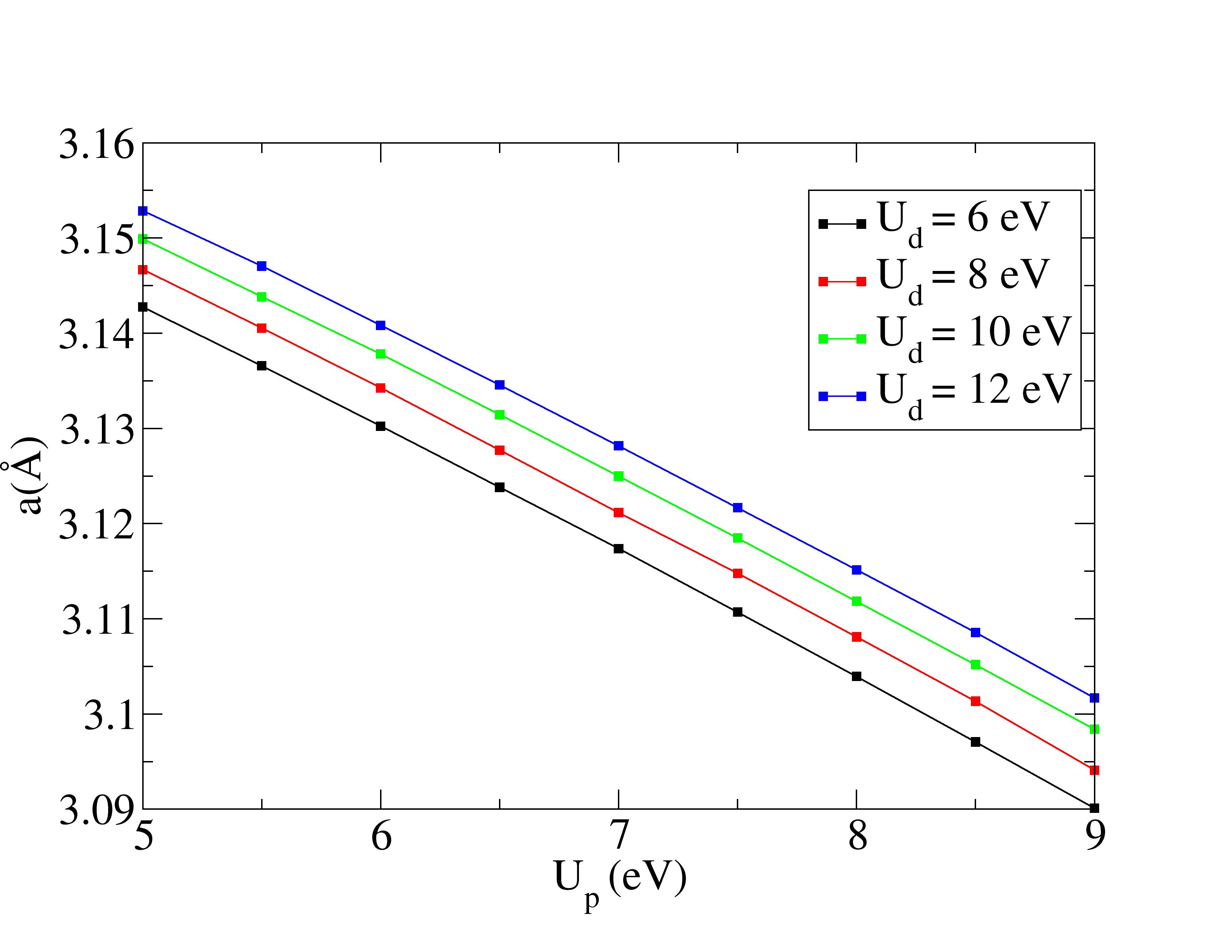} \par}

\centering (a) &
{\centering 
\includegraphics[width=2.7201in,height=2.102in]{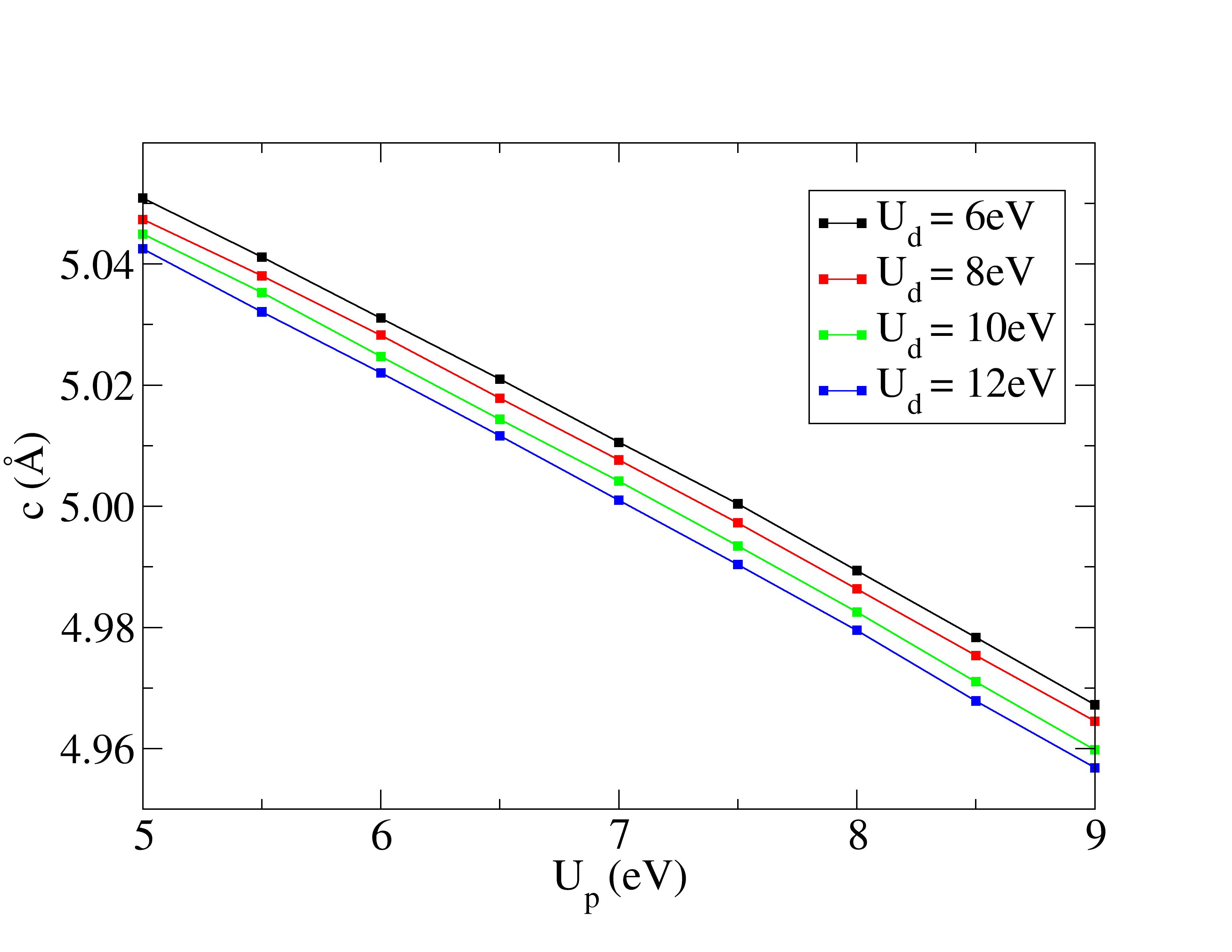} \par}

\centering\arraybslash (b)\\
{\centering 
\includegraphics[width=2.7201in,height=2.102in]{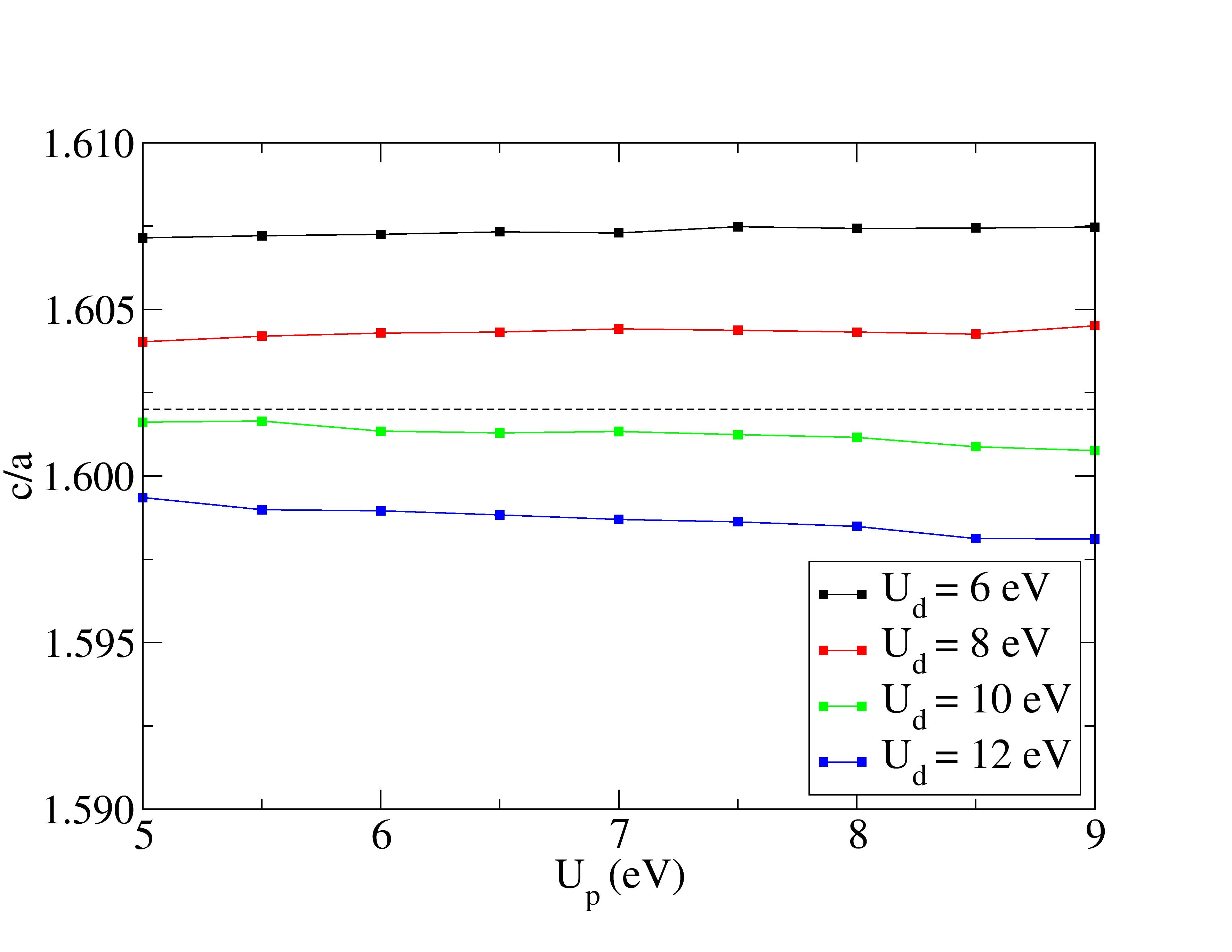} \par}

\centering (c) &
{\centering 
\includegraphics[width=2.7201in,height=2.102in]{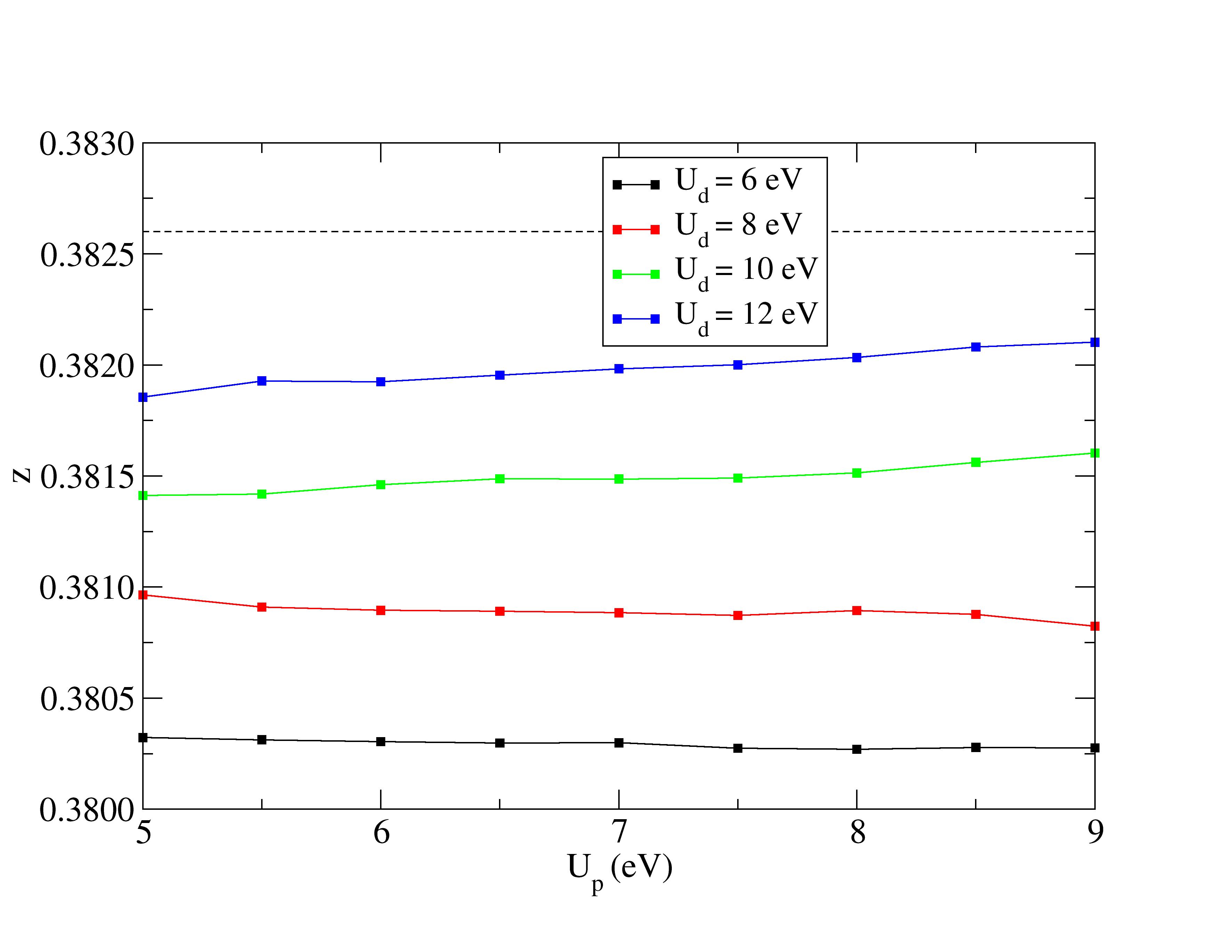} \par}

\centering\arraybslash \textcolor{black}{(d)}\\
\end{supertabular}
\end{flushleft}
\caption{
 Variation of (a) lattice constant $a$, (b) lattice constant $c$, (c) $c/a$ ratio and (d) wurtzite parameter $z$ with the Hubbard term $U_p$ applied to O p-orbitals with fixed $U_d$ values. The dash line corresponds to the corresponding experimental values.
}
\end{figure}

\begin{figure}
\label{fig4}
{\centering 
\includegraphics[width=3.4028in,height=2.6299in]{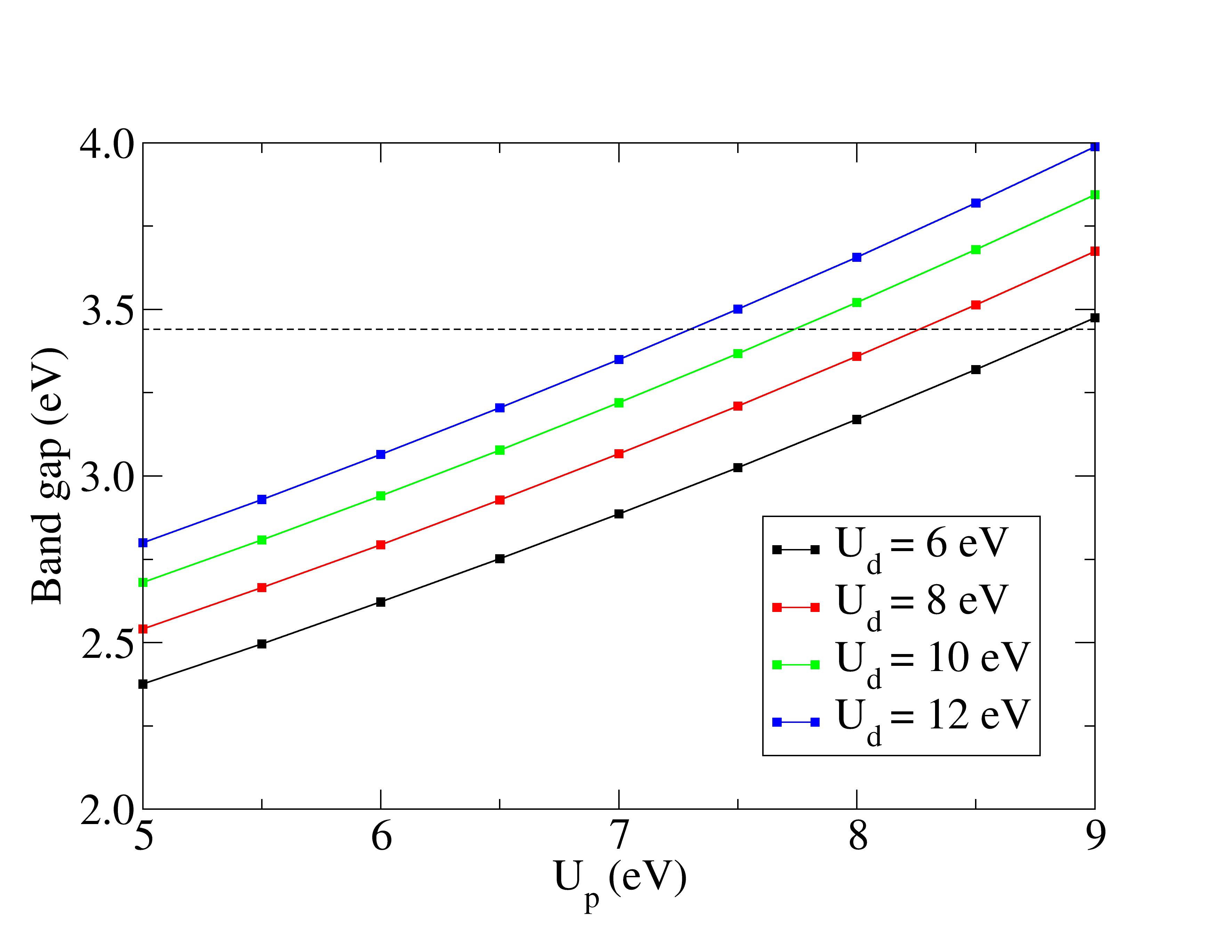} \par}

\caption{Variation of band gap with respect to Hubbard parameter $U_p$ applied to O p-orbitals with fixed $U_d$ values.}
\end{figure}
\bigskip

\begin{figure}[t]
\label{fig5}
{\centering 
\includegraphics[width=3.0528in,height=3.9134in]{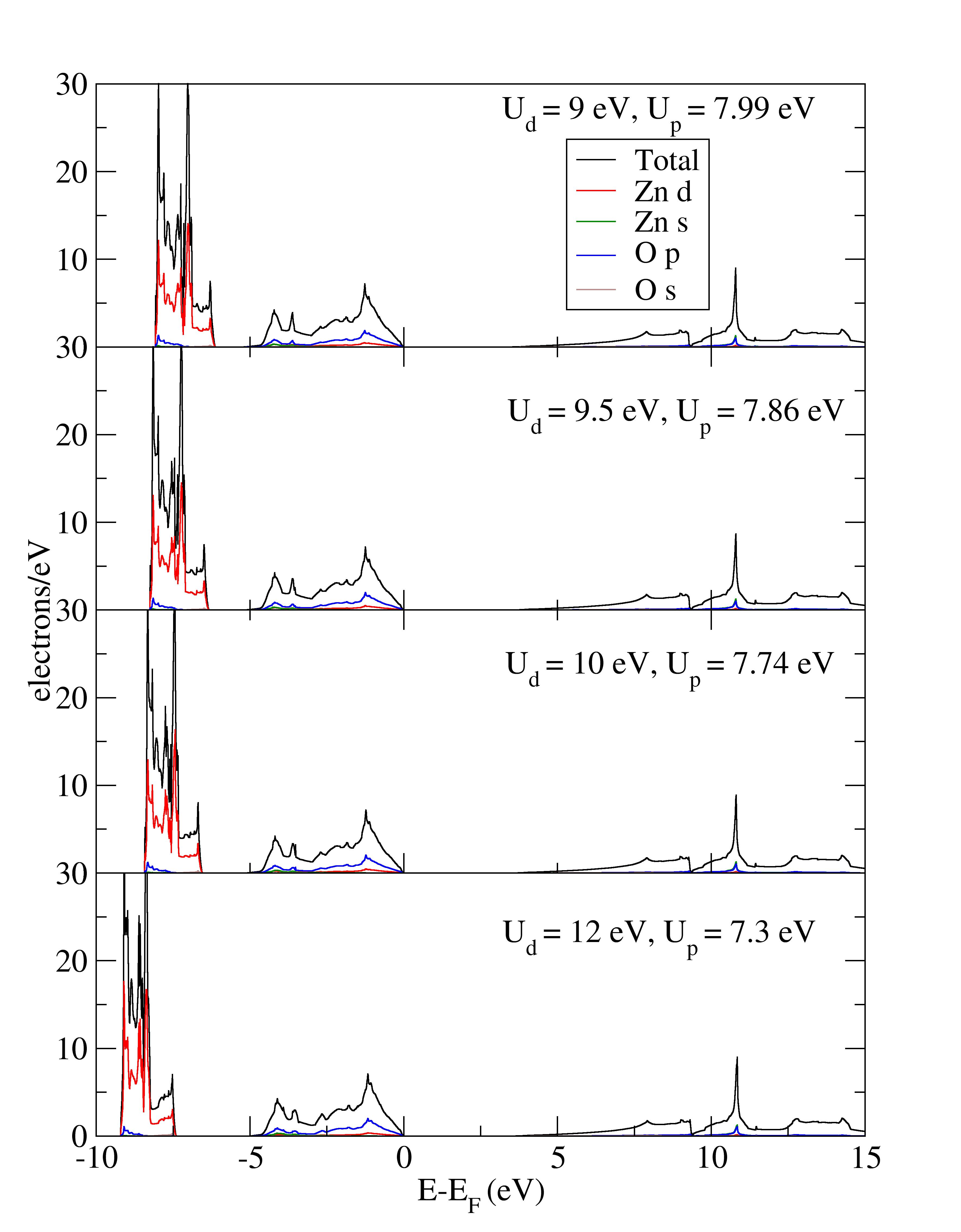} \par}
\caption{Projected density of states (PDOS) diagrams for selected $U_d$ and $U_p$ values matched to the experimental band gap of 3.44 eV.}
\end{figure}

\begin{table}
\label{table3}
\caption{
Estimated structural parameters and electronic band properties based on selected $U_d$ and $U_p$ values (in eV) matched to the experimental band gap of 3.44eV. $E_d$: cation d-band positions. $W_d$: cation d-band bandwidth. $W_p$: anion p valence bandwidth (in eV). The experimental values are referred from Ref [20-22].
}
\begin{center}
\tablehead{}
\begin{supertabular}{m{0.37685984in}m{0.38585985in}m{0.41505986in}m{0.41505986in}m{0.41505986in}m{0.47755983in}m{0.47815987in}m{0.41505986in}m{0.5691598in}m{0.41505986in}m{0.5976598in}}
\hline
\centering $U_d$ &
\centering $U_p$ &
\centering \textit{a }(Å) &
\centering \textit{c }(Å) &
\centering \textit{c/a} &
\centering \textit{z} &
\centering \textit{E}\textit{\textsubscript{d}} &
\centering \textit{W}\textit{\textsubscript{d}} &
\centering Internal gap &
\centering \textit{W}\textit{\textsubscript{p}} &
\centering\arraybslash Valence bandwidth \\\hline
9.0 &
7.99 &
3.110 &
4.985 &
1.603 &
0.3812 &
7.495 &
1.968 &
0.946 &
5.161 &
8.075\\
9.5 &
7.86 &
3.113 &
4.987 &
1.602 &
0.3813 &
7.685 &
1.931 &
1.208 &
5.102 &
8.242\\
10.0 &
7.74 &
3.115 &
4.988 &
1.601 &
0.3815 &
7.875 &
1.888 &
1.387 &
5.137 &
8.412\\
12.0 &
7.30 &
3.124 &
4.995 &
1.600 &
0.3820 &
8.740 &
1.837 &
2.369 &
4.998 &
9.204\\\hline
\multicolumn{2}{m{0.8414598in}}{\centering Experiment} &
3.2427 &
5.1948 &
1.602 &
0.3826 &
7.8 &
2.5 &
1.2 &
5.3 &
9\\\hline
\end{supertabular}
\end{center}
\end{table}

{Analysis of projected density of states (PDOS)
is done to ascertain the modification of the electronic band due to
addition of Hubbard terms. A few sets of
}$U_d${
and
}$U_p${
values are selected for their capability to reproduce band gap value
close to the corresponding experimental value of 3.44 eV. The PDOS
diagrams in }{Figure
5}{ with the corresponding analysis tabulated
in }{Table 3}{ show
that the position of the cationic Zn d-band is heavily dependent on the
selection of the values of Hubbard terms. The addition of Hubbard terms
effectively lowers the position of the Zn 3d states to mitigate the
artificially strong hybridization with O p states provided by
inadequate description in LDA.
}$U_d${=10
eV is shown to be able to produce the d-band energy in good agreement
with experiment whereas
}$U_d${=9
eV produces a shallower d-band energy level and
}$U_d${=12
eV overestimates the magnitude of the d-band energy by 1 eV. All test
cases produced narrower cation d-band bandwidth and anion p valence
bandwidth compared to experiments, particularly for the d-band
bandwidth. The internal band gap }{between the
d states and p states is another quantity sensitive to the change in
the Hubbard terms values, directly affecting the position of the
adjacent d-bands and hence p-d coupling strength. The test case
corresponding to
}$U_d${=9.5
eV and
}$U_p${=7.86
eV produces an accurate internal band gap comparable to the
experimental value 1.2 eV. The valence bandwidths have increased with
the values of
}$U_d${,
even though the cation d-band bandwidth and anion p valence bandwidth
both decrease. This is explained by noticing that the internal band gap
expands with a much higher rate compared to the narrowing of both d and
p bandwidth as
}$U_d${
increases. Taking the d-band energy position and the sensitive internal
band gap as a benchmark, either the pair of
}$U_d${=9.5
eV and
}$U_p${=7.86
eV or
}$U_d${=10.0
eV and
}$U_p${=7.74
eV is the optimum choice for estimation of accurate band gap, which is
close to the choice of Huang }{\textit{et
al.}}{
[}{10}{]. }

{Figure 6}{ shows the
overlap between the band structures obtained using both the standard
LDA and
LDA+}{\textit{U}}{
with selected
}$U_d${=10.0
eV and
}$U_p${=7.74
eV, aligned to the fermi energy level. The conduction band minimum
(CBM) is exactly similar in terms of width and dispersion for both
calculations, differing only by a translation of energy in
LDA+}{\textit{U}}{
due to widening of band gap. The two curves almost overlap at the
valence band maximum (VBM), but slight deviation starts to appear at 4
eV below the fermi energy level. The similar dispersions observed for
both calculations at CBM and VBM indicate the effective mass of
electrons and holes at each region is not affected by the application
of the Hubbard terms. Narrowing of the lower valence bands is observed
for the
LDA+}{\textit{U}}{
results compared to that of LDA, which is consistent with the reduction
in cation d-band bandwidth in the PDOS in
}{Figure 5}{. }

\begin{figure}
\label{fig6}
{\centering 
\includegraphics[width=3.3035in,height=2.5516in]{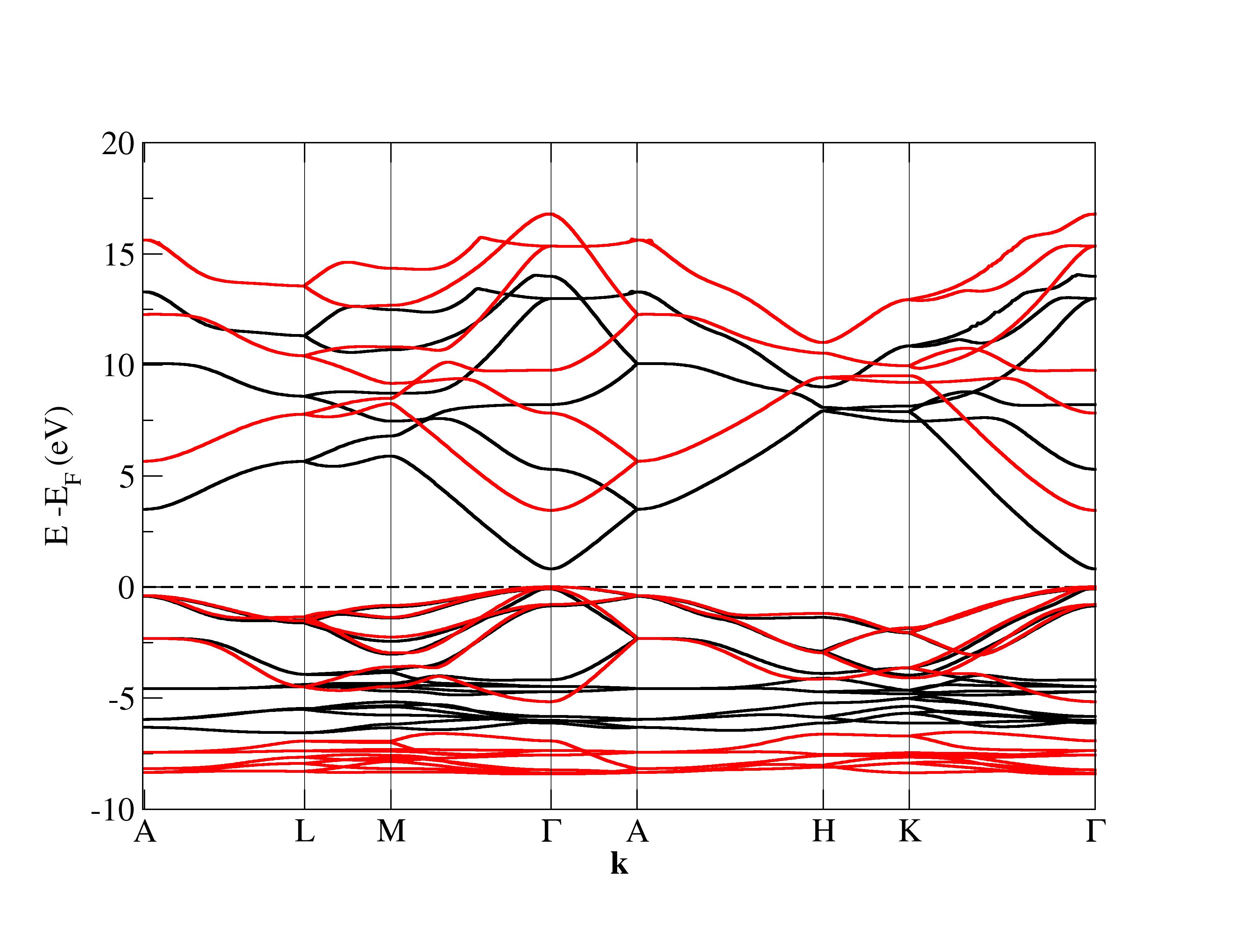} \par}
\caption{Comparisons between the band structures obtained from a standard LDA calculation and a LDA+U calculation with $U_d$=10 eV and $U_p$=7.74 eV.}

\end{figure}


{The investigation is finalized by extending
our work to vibrational properties of wurtzite ZnO at the center of
Brillouin zone with the optimum
}$U_d${=10.0
eV and
}$U_p${=7.74
eV pair. An analysis of the gamma phonon mode
}{Table 4}{ reveals
mismatch between experiment phonon frequencies estimated by
}{LDA+}{\textit{U
}}{using linear response method. Compared to
LDA,
LDA+}{\textit{U}}{
overestimates the gamma point frequencies by about 10\% for the higher
lying modes. It is noticed that gamma phonon modes frequencies in good
agreement with experiment is reproduced by
PBE+}{\textit{U}}{
calculations [}{23}{,
}{24}{]. \ The choice
of change correlation functional plays a huge role in the accurate
description of the underlying electrostatics and vibrational properties
of a system. On the other hand,
LDA+}{\textit{U}}{
correctly reproduces the correct order of gamma centered phonon mode,
in contrast to the higher frequency
B}{\textsubscript{1}}{
mode estimated by LDA.}


\begin{table}
\caption{Phonon mode frequencies (in cm$^{-1}$) at Gamma point of Brillouin zone.}
\label{table4}
\begin{center}
\tablehead{}
\begin{supertabular}{m{0.5150598in}m{0.41505986in}m{0.5518598in}m{0.95665985in}}
\hline
\centering {Mode} &
\centering {LDA} &
\centering
{LDA+}{\textit{U}} &
\centering\arraybslash {Experiment
[}{25}{]}\\\hline
{E}{\textsubscript{2}}
&
{92.5} &
{112.1} &
{98.4}\\
{B}{\textsubscript{1}}
&
{258.4} &
{299.4} &
{258.9}\\
{A}{\textsubscript{1}}{(TO)}
&
{376.5} &
{433.5} &
{378.3}\\
{E}{\textsubscript{1}}{(TO)}
&
{415.8} &
{466.5} &
{412.1}\\
{E}{\textsubscript{2}}
&
{447.9} &
{495.9} &
{438.8}\\
{B}{\textsubscript{1}}
&
{544.7} &
{623.4} &
{551.7}\\
{A}{\textsubscript{1}}{(LO)}
&
{513.6} &
{638.4} &
{573.5}\\
{E}{\textsubscript{1}}{(LO)}
&
{532.6} &
{646.9} &
{592.8}\\\hline
\end{supertabular}
\end{center}
\end{table}

\section{Conclusion}

This study attempts to identify suitable values of Hubbard parameters in
order to produce an accurate band gap estimate for wurtzite ZnO. It is
noticed that the predicted lattice constant \textit{a} and \textit{c}
and wurtzite parameter \textit{z} is low compared to experiment with
the addition of $U_d$ alone. The same is
true for the band gap which is still very low. The addition of Hubbard
term $U_p$ serves to increase the band
gap on par with the magnitude of experimental value, but with the same
underestimation of lattice constants. By constraining the band gap to
that of experiment, the d-band energy value and the bandwidths for
cationic d states and anion p valence states appear to approximate that
of experiment with the {pairs of
}$U_d${=9.5
eV and
}$U_p${=7.86
eV as well as
}$U_d${=10.0
eV and
}$U_p${=7.74
eV, depending on the benchmarking parameters. The internal band gap is
very sensitive to the changes in Hubbard terms, directly affecting the
d-band level and the hybridization between d and p orbitals. However,
the mismatch between predicted gamma point phonon frequencies and the
corresponding experiment values illustrates the need of caution of
applying
LDA+}{\textit{U}}{
calculation to wurtzite ZnO in describing its vibrational properties.}

\section*{Acknowledgement}

This work is supported by Universiti Sains Malaysia RU grant (No.
1001/PFIZIK/811240). We gladfully acknowledge Dr. Chan Huah Yong from
the School of Computer Science, USM, for providing us computing
resources to carry out part of the calculations done in this paper.

\section*{References}

[\label{bkm:BIBanderson2009fundamentals}1]\ \ Anderson Janotti and Chris
G~Van de~Walle. Fundamentals of zinc oxide as a semiconductor.
\textit{Reports on Progress in Physics}, 72 (12): 126501, 2009. URL
\href{http://stacks.iop.org/0034-4885/72/i=12/a=126501}{http://­stacks.iop.org/­0034-4885/­72/­i=12/­a=126501}.

[\label{bkm:BIBperdew1985density}2]\ \ John~P. Perdew. Density
functional theory and the band gap problem. \textit{International
Journal of Quantum Chemistry}, 28 (S19): 497--523, 1985. ISSN
1097-461X. doi:
\href{http://dx.doi.org/10.1002/qua.560280846}{10.1002/qua.560280846}.
URL
\href{http://dx.doi.org/10.1002/qua.560280846}{http://­dx.doi.org/­10.1002/­qua.560280846}.

[\label{bkm:BIBaryasetiawan1998the}3]\ \ F~Aryasetiawan and
O~Gunnarsson. The gw method. \textit{Reports on Progress in Physics},
61 (3): 237, 1998. URL
\href{http://stacks.iop.org/0034-4885/61/i=3/a=002}{http://­stacks.iop.org/­0034-4885/­61/­i=3/­a=002}.

[\label{bkm:BIBzhang2016all}4]\ \ Ming Zhang, Shota Ono, Naoki
Nagatsuka, and Kaoru Ohno. All-electron mixed basis \textit{gw}
calculations of \textit{tio}\textsubscript{2} and zno crystals.
\textit{Phys. Rev. B}, 93: 155116, Apr 2016. doi:
\href{http://dx.doi.org/10.1103/PhysRevB.93.155116}{10.1103/PhysRevB.93.155116}.
URL
\href{http://link.aps.org/doi/10.1103/PhysRevB.93.155116}{http://­link.aps.org/­doi/­10.1103/­PhysRevB.93.155116}.

[\label{bkm:BIBduan2012hybrid}5]\ \ Yifeng Duan, Lixia Qin, Liwei Shi,
Gang Tang, and Hongliang Shi. Hybrid density functional theory study of
band gap tuning in aln and gan through equibiaxial strains.
\textit{Applied Physics Letters}, 100 (2): 022104, 2012. doi:
\href{http://dx.doi.org/10.1063/1.3675864}{10.1063/1.3675864}. URL
\href{http://dx.doi.org/10.1063/1.3675864}{http://­dx.doi.org/­10.1063/­1.3675864}.

[\label{bkm:BIBliechtenstein1995density}6]\ \ A.~I. Liechtenstein, V.~I.
Anisimov, and J.~Zaanen. Density-functional theory and strong
interactions: Orbital ordering in mott-hubbard insulators.
\textit{Phys. Rev. B}, 52: R5467--R5470, Aug 1995. doi:
\href{http://dx.doi.org/10.1103/PhysRevB.52.R5467}{10.1103/PhysRevB.52.R5467}.
URL
\href{http://link.aps.org/doi/10.1103/PhysRevB.52.R5467}{http://­link.aps.org/­doi/­10.1103/­PhysRevB.52.R5467}.

[\label{bkm:BIBdudarev1998electron}7]\ \ S.~L. Dudarev, G.~A. Botton,
S.~Y. Savrasov, C.~J. Humphreys, and A.~P. Sutton. Electron-energy-loss
spectra and the structural stability of nickel oxide: An lsda+u study.
\textit{Phys. Rev. B}, 57: 1505--1509, Jan 1998. doi:
\href{http://dx.doi.org/10.1103/PhysRevB.57.1505}{10.1103/PhysRevB.57.1505}.
URL
\href{http://link.aps.org/doi/10.1103/PhysRevB.57.1505}{http://­link.aps.org/­doi/­10.1103/­PhysRevB.57.1505}.

[\label{bkm:BIBcococcioni2005linear}8]\ \ Matteo Cococcioni and Stefano
de~Gironcoli. Linear response approach to the calculation of the
effective interaction parameters in the \textit{LDA}+\textit{U} method.
\textit{Phys. Rev. B}, 71: 035105, Jan 2005. doi:
\href{http://dx.doi.org/10.1103/PhysRevB.71.035105}{10.1103/PhysRevB.71.035105}.
URL
\href{http://link.aps.org/doi/10.1103/PhysRevB.71.035105}{http://­link.aps.org/­doi/­10.1103/­PhysRevB.71.035105}.

[\label{bkm:BIBagapito2015reformulation}9]\ \ Luis~A. Agapito, Stefano
Curtarolo, and Marco Buongiorno~Nardelli. Reformulation of
\textit{DFT}+\textit{u} as a pseudohybrid hubbard density functional
for accelerated materials discovery. \textit{Phys. Rev. X}, 5: 011006,
Jan 2015. doi:
\href{http://dx.doi.org/10.1103/PhysRevX.5.011006}{10.1103/PhysRevX.5.011006}.
URL
\href{http://link.aps.org/doi/10.1103/PhysRevX.5.011006}{http://­link.aps.org/­doi/­10.1103/­PhysRevX.5.011006}.

[\label{bkm:BIBhuang2012detailed}10]\ \ Gui-Yang Huang, Chong-Yu Wang,
and Jian-Tao Wang. Detailed check of the lda+ u and gga+ u corrected
method for defect calculations in wurtzite zno. \textit{Computer
Physics Communications}, 183 (8): 1749--1752, 2012.

[\label{bkm:BIBsabine1969the}11]\ \ T.~M. Sabine and S.~Hogg. The
wurtzite \textit{Z} parameter for beryllium oxide and zinc oxide.
\textit{Acta Crystallographica Section B}, 25 (11): 2254--2256, Nov
1969. doi:
\href{http://dx.doi.org/10.1107/S0567740869005528}{10.1107/S0567740869005528}.
URL
\href{https://doi.org/10.1107/S0567740869005528}{https://­doi.org/­10.1107/­S0567740869005528}.

[\label{bkm:BIBgonze2009abinit}12]\ \ X.~Gonze, B.~Amadon, P.-M.
Anglade, J.-M. Beuken, F.~Bottin, P.~Boulanger, F.~Bruneval,
D.~Caliste, R.~Caracas, M.~Côté, T.~Deutsch, L.~Genovese, Ph. Ghosez,
M.~Giantomassi, S.~Goedecker, D.R. Hamann, P.~Hermet, F.~Jollet,
G.~Jomard, S.~Leroux, M.~Mancini, S.~Mazevet, M.J.T. Oliveira,
G.~Onida, Y.~Pouillon, T.~Rangel, G.-M. Rignanese, D.~Sangalli,
R.~Shaltaf, M.~Torrent, M.J. Verstraete, G.~Zerah, and J.W. Zwanziger.
Abinit: First-principles approach to material and nanosystem
properties. \textit{Computer Physics Communications}, 180 (12): 2582 --
2615, 2009. ISSN 0010-4655. doi:
\href{http://dx.doi.org/10.1016/j.cpc.2009.07.007}{10.1016/j.cpc.2009.07.007}.
URL
\href{http://www.sciencedirect.com/science/article/pii/S0010465509002276}{http://­www.sciencedirect.com/­science/­article/­pii/­S0010465509002276}.

[\label{bkm:BIBblochl1994projector}13]\ \ P.~E. Blöchl. Projector
augmented-wave method. \textit{Phys. Rev. B}, 50: 17953--17979, Dec
1994. doi:
\href{http://dx.doi.org/10.1103/PhysRevB.50.17953}{10.1103/PhysRevB.50.17953}.
URL
\href{http://link.aps.org/doi/10.1103/PhysRevB.50.17953}{http://­link.aps.org/­doi/­10.1103/­PhysRevB.50.17953}.

[\label{bkm:BIBjollet2014generation}14]\ \ François Jollet, Marc
Torrent, and Natalie Holzwarth. Generation of projector augmented-wave
atomic data: A 71 element validated table in the \{XML\} format.
\textit{Computer Physics Communications}, 185 (4): 1246 -- 1254, 2014.
ISSN 0010-4655. doi:
\href{http://dx.doi.org/10.1016/j.cpc.2013.12.023}{10.1016/j.cpc.2013.12.023}.
URL
\href{http://www.sciencedirect.com/science/article/pii/S0010465513004359}{http://­www.sciencedirect.com/­science/­article/­pii/­S0010465513004359}.

[\label{bkm:BIBamadon2008gamma}15]\ \ B.~Amadon, F.~Jollet, and
M.~Torrent. $\gamma $ and $\beta $ cerium: Lda+ u calculations of
ground-state parameters. \textit{Phys. Rev. B}, 77: 155104, Apr 2008.
doi:
\href{http://dx.doi.org/10.1103/PhysRevB.77.155104}{10.1103/PhysRevB.77.155104}.
URL
\href{http://link.aps.org/doi/10.1103/PhysRevB.77.155104}{http://­link.aps.org/­doi/­10.1103/­PhysRevB.77.155104}.

[\label{bkm:BIBmang1995band}16]\ \ A~Mang, K~Reimann, et~al. Band gaps,
crystal-field splitting, spin-orbit coupling, and exciton binding
energies in zno under hydrostatic pressure. \textit{Solid State
Communications}, 94 (4): 251--254, 1995.

[\label{bkm:BIBwalle1999correcting}17]\ \ A.~van~de Walle and G.~Ceder.
Correcting overbinding in local-density-approximation calculations.
\textit{Phys. Rev. B}, 59: 14992--15001, Jun 1999. doi:
\href{http://dx.doi.org/10.1103/PhysRevB.59.14992}{10.1103/PhysRevB.59.14992}.
URL
\href{http://link.aps.org/doi/10.1103/PhysRevB.59.14992}{http://­link.aps.org/­doi/­10.1103/­PhysRevB.59.14992}.

[\label{bkm:BIBkaczkowski2012electronic}18]\ \ J~Kaczkowski. Electronic
structure of some wurtzite semiconductors: hybrid functionals vs. ab
initio many body calculations. \textit{Acta Physica Polonica A}, 121:
1142--1144, 2012. doi:
\href{http://dx.doi.org/10.12693/APhysPolA.121.1142}{10.12693/APhysPolA.121.1142}.

[\label{bkm:BIBma2013correlation}19]\ \ Xinguo Ma, Ying Wu, Yanhui Lv,
and Yongfa Zhu. Correlation effects on lattice relaxation and
electronic structure of zno within the gga+u formalism. \textit{The
Journal of Physical Chemistry C}, 117 (49): 26029--26039, 2013. doi:
\href{http://dx.doi.org/10.1021/jp407281x}{10.1021/jp407281x}. URL
\href{http://dx.doi.org/10.1021/jp407281x}{http://­dx.doi.org/­10.1021/­jp407281x}.

[\label{bkm:BIBozgur2005a}20]\ \ Ü. Özgür, Ya.~I. Alivov, C.~Liu,
A.~Teke, M.~A. Reshchikov, S.~Dog\u{a}n, V.~Avrutin, S.-J. Cho, and
H.~Morkoç. A comprehensive review of zno materials and devices.
\textit{Journal of Applied Physics}, 98 (4): 041301, 2005. doi:
\href{http://dx.doi.org/10.1063/1.1992666}{10.1063/1.1992666}. URL
\href{http://dx.doi.org/10.1063/1.1992666}{http://­dx.doi.org/­10.1063/­1.1992666}.

[21\label{bkm:B4Bwaroquiers2013band}]\ \ David Waroquiers, Aurélien
Lherbier, Anna Miglio, Martin Stankovski, Samuel Poncé, Micael J.~T.
Oliveira, Matteo Giantomassi, Gian-Marco Rignanese, and Xavier Gonze.
Band widths and gaps from the tran-blaha functional: Comparison with
many-body perturbation theory. \textit{Phys. Rev. B}, 87: 075121, Feb
2013. doi:
\href{http://dx.doi.org/10.1103/PhysRevB.87.075121}{10.1103/PhysRevB.87.075121}.
URL
\href{http://link.aps.org/doi/10.1103/PhysRevB.87.075121}{http://­link.aps.org/­doi/­10.1103/­PhysRevB.87.075121}.

[\label{bkm:BIBvogel1995ab}22]\ \ Dirk Vogel, Peter Krüger, and Johannes
Pollmann. Ab initio. \textit{Phys. Rev. B}, 52: R14316--R14319, Nov
1995. doi:
\href{http://dx.doi.org/10.1103/PhysRevB.52.R14316}{10.1103/PhysRevB.52.R14316}.
URL
\href{http://link.aps.org/doi/10.1103/PhysRevB.52.R14316}{http://­link.aps.org/­doi/­10.1103/­PhysRevB.52.R14316}.

[\label{bkm:BIBcalzolari2013dielectric}23]\ \ Arrigo Calzolari and
Marco~Buongiorno Nardelli. Dielectric properties and raman spectra of
zno from a first principles finite-differences/finite-fields approach.
\textit{Scientific reports}, 3, 2013.

[\label{bkm:BIBconsiglio2016importance}24]\ \ Anthony Consiglio and
Zhiting Tian. Importance of the hubbard correction on the thermal
conductivity calculation of strongly correlated materials: a case study
of zno. \textit{Scientific Reports}, 6, 2016.

[\label{bkm:BIBserrano2010phonon}25]\ \ J.~Serrano, F.~J. Manjón, A.~H.
Romero, A.~Ivanov, M.~Cardona, R.~Lauck, A.~Bosak, and M.~Krisch.
Phonon dispersion relations of zinc oxide: Inelastic neutron scattering
and ab initio calculations. \textit{Phys. Rev. B}, 81: 174304, May
2010. doi:
\href{http://dx.doi.org/10.1103/PhysRevB.81.174304}{10.1103/PhysRevB.81.174304}.
URL
\href{http://link.aps.org/doi/10.1103/PhysRevB.81.174304}{http://­link.aps.org/­doi/­10.1103/­PhysRevB.81.174304}.

\bigskip
\end{document}